# Interactive Digital Learning Materials for Kindergarten Students in Bangladesh

Md. Baharul Islam[1], Dr. Md. Kabirul Islam[1], Arif Ahmed[2], Abu Kalam Shamsuddin[1]

[1]*Department of Multimedia Technology and Creative Arts, Daffodil International University, Dhaka-1207, Bangladesh*

[2]*AAVA 3D, Mirpur 12, Dhaka 1216, Bangladesh*

baharul@daffodilvarsity.edu.bd
kislam@daffodilvarsity.edu.bd
aava3dfx@gmail.com
shamsuddin@daffodilvarsity.edu.bd

*Abstract*— **Traditional education system for preschool children is not updated in Bangladesh. Generally, parents and teachers are tried to teach children by introducing first alphabet and numbers in the form of text books. Sometimes it is quite difficult for teachers to teach play group about their first learning when they are not interested with it. The pedagogy of teaching and learning is changed with the proliferation of communication technology and it is necessary to develop interactive learning materials for children that may improve their learning, catching, and memorizing capabilities. Perhaps, one of the most important innovations in the age of technology is multimedia and its application. It is imperative to create high quality and realistic leaning environment for children. Interactive learning materials can be easier to understand and deal with their first learning. We developed some interactive learning materials in the form of video for playgroup using multimedia application tools. This study investigated the impact of student's abilities to acquire new knowledge or skills through interactive learning materials. We visited one kindergartens (Nursery schools), interviewed class teachers about their teaching methods and level of students' ability of recognizing English alphabets, pictures etc. The course teachers were provided interactive learning materials to show their playgroups for a number of sessions. The video included English alphabets with related words and pictures, and motivational funs. We noticed that almost all children were very interested to interact with their leaning video. The students were assesses individually and asked to recognize the alphabets, and pictures. The students adapted with their first alphabets very quickly. However, there were individual differences in their cognitive development. This interactive multimedia can be an alternative to traditional pedagogy for teaching playgroups.**

## I. INTRODUCTION

Interactive digital learning materials refers to educational products on compute-based systems which response to the students' actions by presenting contents such as texts, graphics, animation, video, audio etc [1]. Within this learning context, interaction takes place between the students and the contents of the digital system, and eventually learning may occur through interaction. It is perceived that in Bangladesh numerous kindergarten schools impose children to read lots of books at their early stage and memorize the contents when asked to do so. In fact, the beginners (playgroups) do not have any choice for their first learning apart from obeying the instructions of their teachers, house tutors, and parents. The parents invest lot of time for teaching their kids at home and their main objective is to obtain a higher position or grade by their loving son/daughter. Unfortunately, the strategies of learning by interacting with educational aids, learning by playing and learning by doing are ignored in the above teaching style.

The most exciting invention of multimedia technology has created an opportunity to shift from traditional reading and memorization habit to learning by interacting with interesting contents. The internet and other communication media have promoted implementation of this technology. The rapid expansion of multimedia technologies over the last decade has brought about indispensable changes to computing, entertainment, and education regardless of grade levels [2]. We have experience of children's curiosity about computer, mobile phone; play station portable (PSP) games etc. Previous researchers found that kindergarten students gained learning skills through interacting with multimedia contents [3]. Moreover, multimedia has the potential to create high quality learning environments especially for children, with the capability of creating a more pragmatic learning context through its different medias- texts, graphics, sound, animation etc. In Bangladesh use of this technology in kindergarten school is sporadic and no research has been conducted so far into this area. Main purpose of this study was to investigate playgroup students' responses on multimedia contents in learning English alphabets with animated pictures.

## II. LITERATURE REVIEW

By Interactive multimedia, educators unusually refer to the using of multimedia and Information Communication Technology (ICT) equipments are to offer an effective dialog between the resource materials- indirectly with the instructor and the students in comparison with traditional methods of teaching which may lack such interactivity [3]. Modern education and communication environments can offer alternative ways in the learning process. Multimedia has been widely used in educational technologies. It is also expected that future will see more of the utilization of such tools in education. Using interactive multimedia in the teaching process is growing in the present context. Multimedia plays a very important role in assisting students in learning processes [3].

Nusir et.al [3] investigated the possibility of enhancing the early education system with multimedia technologies previously developed to teach students at young ages basic skills. They found the positive impact of the developed program on students' abilities to understand new knowledge or skills. The researchers commented that multimedia education offers an alternative to traditional education that can enhance the current methods and provide an alternative especially in some cases where teaching in educational methods is not applicable. Similarly, another group of researchers [4] developed an English short play as a teaching material to promote children's (second language learners) English learning attitude and interests and was presented to all classmates and evaluated by three professors. The findings of the study reveal that incorporating project-based learning into the development of an English short play can effectively guide students in creating the short play effectively. Additionally, a Chinese folk story based English short play enhanced elementary school students' English learning interests and motivation. Another study [5] sought to explore the effects of contextual cues and support requirements of multimedia animation on children's English learning. Support requirements for design were put into two categories: no support requirements vs. support through display of key images before listening and English description and the display of key images after listening as well as printing materials. The study found: 1) contextual cues play an extremely important role in the process of children's learning via multimedia animation, which means children are more dependent on contextual cues; 2) the design of support requirements needs to provide or complement relevant and specific contextual cues in order to help children's comprehension and match the audio to the context.

An animation media can help children expand their English vocabulary and receive higher average score than those who apply the normal one at statistical significance level of 0.01[6]. The researchers commented that the animation method is a beneficial teaching material to stimulate and support the learners, especially at 5 to 6 years old to enjoy the class with good results. Vate-U-Lan [7] conducted a study in Bangkok where 3D pop-up book was used as a tool for teachers to deliver the story of a children's book to teach various English language aspects to young children and found the resources effective to enhance learning and increase desire to learn. Phonetic awareness is a critical and often neglected component in learning English language especially for children. However, multimedia resources in English were found to significantly improve not only their (children's) English pronunciation but their spelling and reading abilities although there were individual differences [8]. Several studies conducted at Sri Lankan universities in integrating web-based learning and interactive computer aided language learning in English as a Second Language and foreign languages proved to be most successful in enhancing student performance, broadening world knowledge, increasing interest and motivation in autonomous learning. This promising multimedia technology is worthy for both school and higher education [10]. Developing an interactive multimedia courseware with storytelling approach promotes learning process become more exciting and the knowledge could be delivered and acknowledged easier [11]. The courseware was used for 5-6 year old preschool children in Malaysia for promoting and enhancing preschool education in promoting for the development of excellent individual in a knowledge society. It is obvious that method of interaction design for enhancing children's cognitive ability is crucial [12]. The courseware development may include many crucial decision-making stages such as analysis, design, development, implementation and evaluation (ADDIE model) [13].

The literature review indicates that the multimedia courseware developed for teaching children especially preschool students has a positive impact for promoting learning but designing and developing the materials for their engagement are essential aspects for paying careful attention.

## III. METHODOLOGY

The core objectives of this study were to find out playgroup students' responses to an interactive digital material (video) on English alphabets and to identify children's abilities to cope with the digital learning materials. We have completed our study in three phases.

Firstly, we have developed 26 English alphabets with correspond words and objects like A for Apple, B for Ball, C for Car etc as frame by frame. We developed not only words but also corresponding objects using different multimedia software and applications. How to write the each alphabet also included in our works. We tried to give British accent on our learning materials so that students can learn original pronunciation of different alphabets and words. We also developed 10 Bengali digits (1 to 10) and corresponding objects so that children can easily recognize and adapt with the digits interestingly. Then we had to rearrange all frames and produced it in the form of interactive learning video for children.

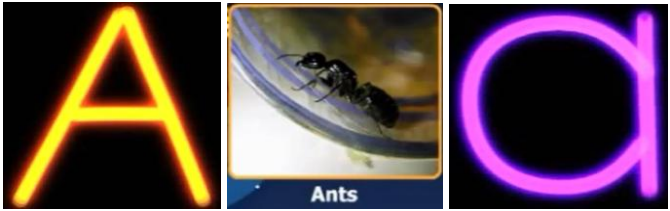
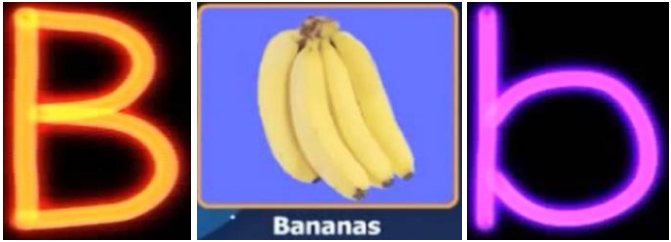
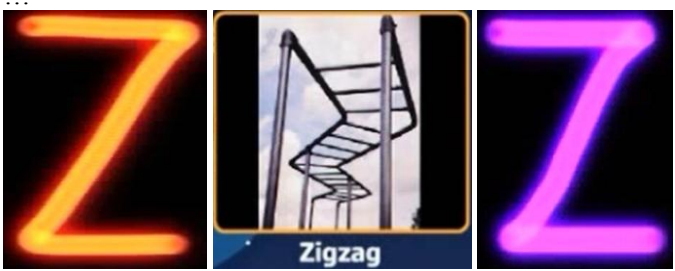

Fig. 1 Developed alphabet with corresponding words and image in form of video

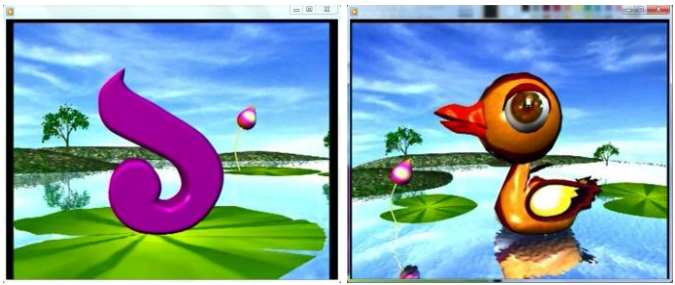
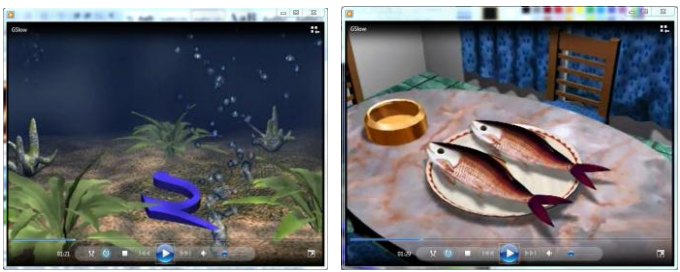

Fig 2 Developed Bengali digit with corresponding objects in form of video

Secondly, we visited one kindergarten school named Ahsania Mohila Mission High School in Dhaka. There were 52 kindergarten children (aged 4-5 years). We took a pre-test before showing our developed learning materials through the help of class teachers. Almost all children already are known the alphabet but only few children can make word using each alphabet.

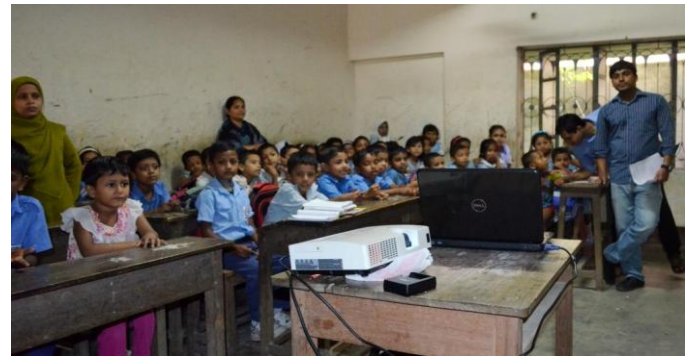

Fig. 3 playing interactive learning materials in front of kindergarten students in Ahsania Mohila Mission High School

In the phase of third, we showed our developed interactive learning materials using projector. In figure 3 we demonstrate our learning materials in Ahsania Mohila Mission High School where 52 kindergarten students. And we played video one time and asked the children individually about alphabet with correspond words and objects. Most of the children recognized the words and objects. Only few students could not identify objects with corresponding words. For them, we played the same video again and asked the weak children second time. Surprisingly, they recognized objects and words more than the first time. Possibly they got some ideas from first time video display and received help from other successful children.

## IV. RESULT AND DISCUSSION

The Table 1 and Figure 3 show the results of pre-test before playing interactive learning materials. There were 19 students already know the alphabets, digits, words with corresponding objects because of test held at the end of academic year. They knew it from the home and school said their class teacher.

TABLE I: PRE-TEST RESULT FOR STUDENTS

| No. of Students | Recognized Alphabets | Recognized words with objects | Recognized Digits |
|---|---|---|---|
| 19 | Yes | Yes | Yes |
| 12 | Yes | No | Yes |
| 11 | Yes | No | No |
| 6 | No | No | Yes |
| 4 | No | No | No |
| TOTAL: 52 | 46 | 19 | 37 |

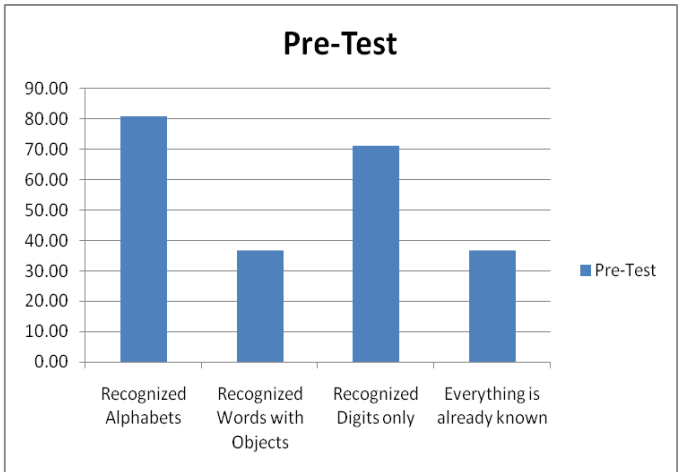
Fig 3 Pre-test result as Bar chart

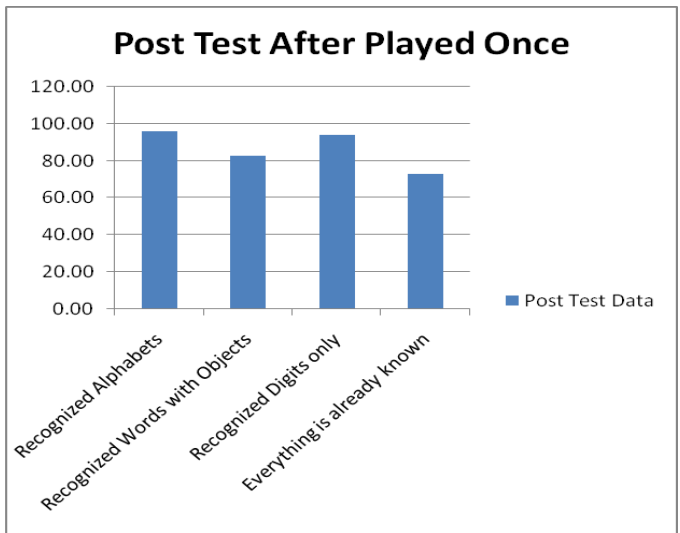
Fig. 4 Post test result after played interactive learning materials once only

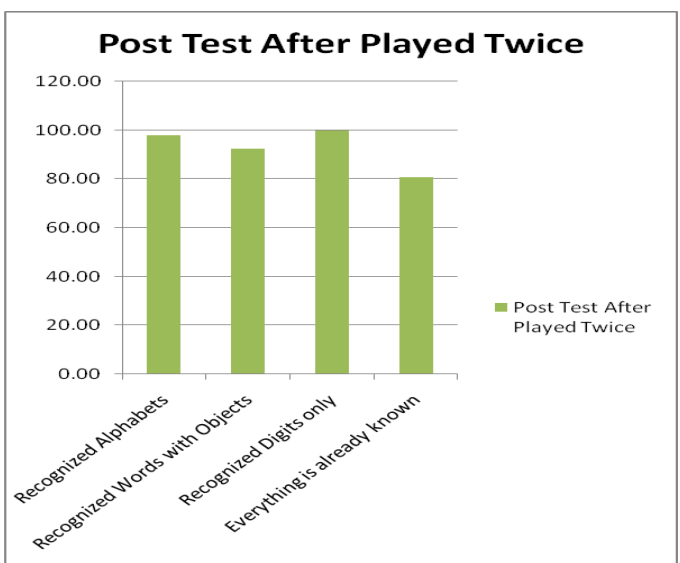
Fig. 5 Post test result after played learning materials twice

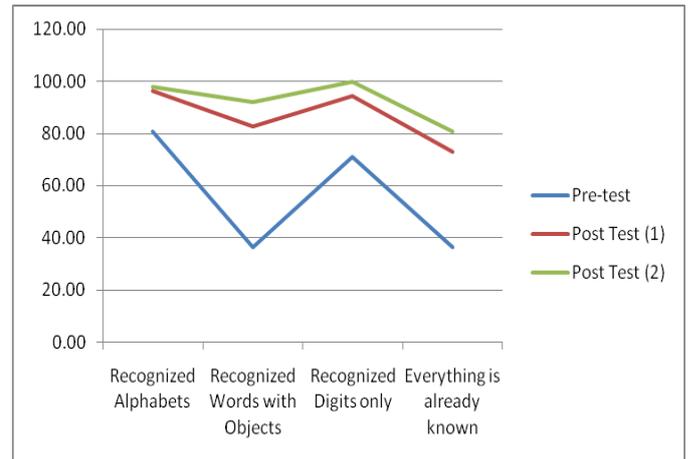
Fig. 6 Pre and Post test result of our study

Figure 4, 5 and 6 show the results of our interactive learning materials after played zero, once and twice respectively. The graph indicates that there were individual differences in students' responses toward the interactive learning video [8]. Students are recognized alphabets, digits and words with objects more and more according to the number played materials. Almost all students learned after played twice the learning materials. We asked Ms Sirajia Begum about our developed interactive learning materials who is working as head of the school. She said,

*"Definitely these materials will be helpful for pre-school and kindergarten students. They will be quick learner with enjoy and fun."*

We also wanted to know the feeling of students when learning materials was playing. All students were very enthusiastic and wanted to see the video more we stopped playing our video. All students pronounced loudly when a most recognizable alphabet with object came. They learned how to write alphabet quickly after watching the learning materials. It indicates that they had sufficient ability to adapt with the learning process [3], [6].

## V. CONCLUTIONS AND FUTUTE WORKS

In this study, we developed two interactive learning materials for kindergarten and play level school children in Bangladesh. One learning materials was developed based on English alphabet with the corresponding words and objects where another materials was developed based on Bengali digit from 0 to 10. We used different types of multimedia applications and softwares to develop these materials. We are noticed that the impact of interactive learning materials is exclusively high to improve their learning skills and adaptation. Although our method is showing the improvement of kindergarten children learning skills, we have no intention the replacement of traditional education system. Just we wanted to show a method to enhance the skill and quick adaptation with their learning. In future, we will develop a writing method so that children can improve their writing skills through interactive learning materials. We will do it not only for Bangladeshi children but also the globe. We will assess the effectiveness of learning materials through experiment of large groups. We have to measure the weakness of our proposed system. We hope that this research will be helpful for those who want to work with interactive learning methods for children.

## ACKNOWLEDGEMENT
We are thanking to Ms. Sirajia Begum and Ms. Ratna Pervin, Teacher of Ahsania Mohila Mission High School for their excellent cooperation for testing and sharing our materials with the children of the school. We also thank to Ms. Ambia Ferdoys and Mr. Khondokar Hasibul Islam for their helps in different ways.